\documentclass{article}[12pt]
\usepackage[latin1]{inputenc}
\usepackage{graphics,graphicx,longtable,verbatim}
\usepackage{setspace}
\usepackage{amsmath}
\usepackage{amssymb}
\usepackage{bbm}
\usepackage[english]{babel}
\usepackage{hyperref,float}
\usepackage[authoryear]{natbib}
\usepackage[top=1.5in, bottom=1.5in, left=1in, right=1in]{geometry}
\usepackage[]{algorithm2e}

\title{Directional Assessment of Traffic Flow Extremes}
\author{Maria Osipenko \\ Department of Business and Economics, Berlin School of Economics and Law, Germany}
\date{}

\def\argmin{\mathop{\textup{argmin}}}
\def\argmax{\mathop{\textup{argmax}}}
\newcommand{\rE}{\text{\sffamily{E}}}

\begin{document}
\maketitle
\begin{abstract}
We analyze extremes of traffic flow profiles composed of traffic counts over a day. The data is essentially curves and determining which trajectory should be classified as extreme is not straight forward. 
To assess the extremes of the traffic flow curves in a coherent way, we use a directional definition of extremeness and apply the dimension reduction technique called  principal component analysis (PCA) in an asymmetric norm.
In the classical PCA one reduces the dimensions of the data by projecting it in the direction of the largest variation of the projection around its mean. In the PCA in an asymmetric norm one chooses the projection directions, such that the asymmetrically weighted variation around a tail index -- an expectile -- of the data  is the largest possible. Expectiles are tail measures that generalize the mean in a similar manner as quantiles generalize the median. Focusing on the asymmetrically weighted variation around an expectile of the data, we find the appropriate projection directions and the low dimensional representation of the traffic flow profiles that uncover different patterns in their extremes. 
Using the traffic flow data from the roundabout on Ernst-Reuter-Platz in the city center of Berlin, Germany, we estimate, visualize and interpret the resulting principal expectile components. 
The corresponding directional extremes of the traffic flow profiles are simple to identify and to connect to their location- and time-related specifics. Their shapes are driven by their scores on each principal expectile component which is useful for extracting and analyzing traffic patterns. 
Our approach to dimensionality reduction towards  the directional extremes of traffic flow extends the related methodological basis and gives promising results for subsequent analysis, prediction and control of traffic flow patterns.
\end{abstract}

%%%%%%%%%%%%%%%%%%%%%%%%%%%%%%%%%%%%%%%%
\section{Background}
Intelligent traffic planing in smart city context (e.g.\cite{Neirotti}) is of  immense importance for big cities, where the available traffic infrastructure is already at the limit. It has the potential to meet mobility needs while reducing the environmental impact of traffic and increasing life quality, as \cite{Julita}. Such planing requires the identification of  long term traffic patterns.

Exploring day-to-day variability in the shapes of daily traffic profiles, composed of traffic measurements over a day, is an important part in the planing process. Various strategies for analyzing such profile curves from visual inspection e.g. in \cite{SongYing} to different clustering methods e.g. in \cite{Fran}, \cite{Yang} have been proposed. 
Especially appealing are the methods resulting in a low dimensional representation of such traffic patterns which enhances discovering of useful structures, data visualization, predictive modeling and network analysis as pointed out in \cite{Jin}, \cite{MIN2011606}, \cite{prediction} and \cite{BATTERMAN2015}.

\cite{KAYANI} successfully apply partitioning around medoids to shape characteristics of traffic flow. \cite{Yang} make use of spectral clustering to group average speed profiles based on their similarity graphs.  \cite{Naranjo} references further clustering applications. The resulting lower dimensional representation is the cluster membership used to interpret the clusters and to conduct subsequent modeling and prediction. 

\cite{tsas}, \cite{Xing} and \cite{song2019} among others show that road traffic data is well represented as a mixture of persistent underlying patterns and the well known principal component analysis (PCA) gives easily obtainable and interpretable insights to the traffic structure in terms of the variability around the mean.

The classical PCA  (\cite{pca}) and its functional and robust versions have been successfully applied to model travel time (\cite{ZHONG}) and flow profiles ( \cite{GUARDIOLA2014119}, \cite{Xing}, \cite{COOGAN}, and \cite{CRAWFORD2017}). These approaches seek to project the data in the direction of the largest variation around its mean which might be sub-optimal when one is interested in the extremes of the data. In particular, when dealing with observations originating from skewed distributions or from mixtures of distributions, shifting the focus to some tail index and considering  variations around this tail index may uncover interesting structures in the data extremes.  As outlined in \cite{Yu} and \cite{COOGAN}, capturing those relevant features in the behavior of traffic extremes may enhance ongoing applications in traffic forecasting and control.

In order to analyze extremes of traffic data, we need to have a sense of extremeness. The classical extremeness measures for univariate data are quantiles (\cite{koenker}). For a real valued random variable $X$ with the cumulative distribution function $F_X$ and $0<\tau<1$, the $\tau$-quantile $q_{\tau}$ corresponds to the inverse of the cumulative distribution function of $X$ in $\tau$, formally  $q_{\tau}= F_X^{-1}(\tau)$.

A tail measure similar to the $\tau$-quantile is the $\tau$-expectile introduced in \cite{newey}. The $\tau$-expectile is the real number, such that the $\tau$-proportion of the expected distance to it lies below and the $(1-\tau)$-proportion lies above it.
As quantiles generalize the concept of the median, expectiles generalize the concept of the mean.  

\begin{figure}[hb!]
	\begin{center}
		\label{intro_lines}
\includegraphics[width=0.65\textheight]{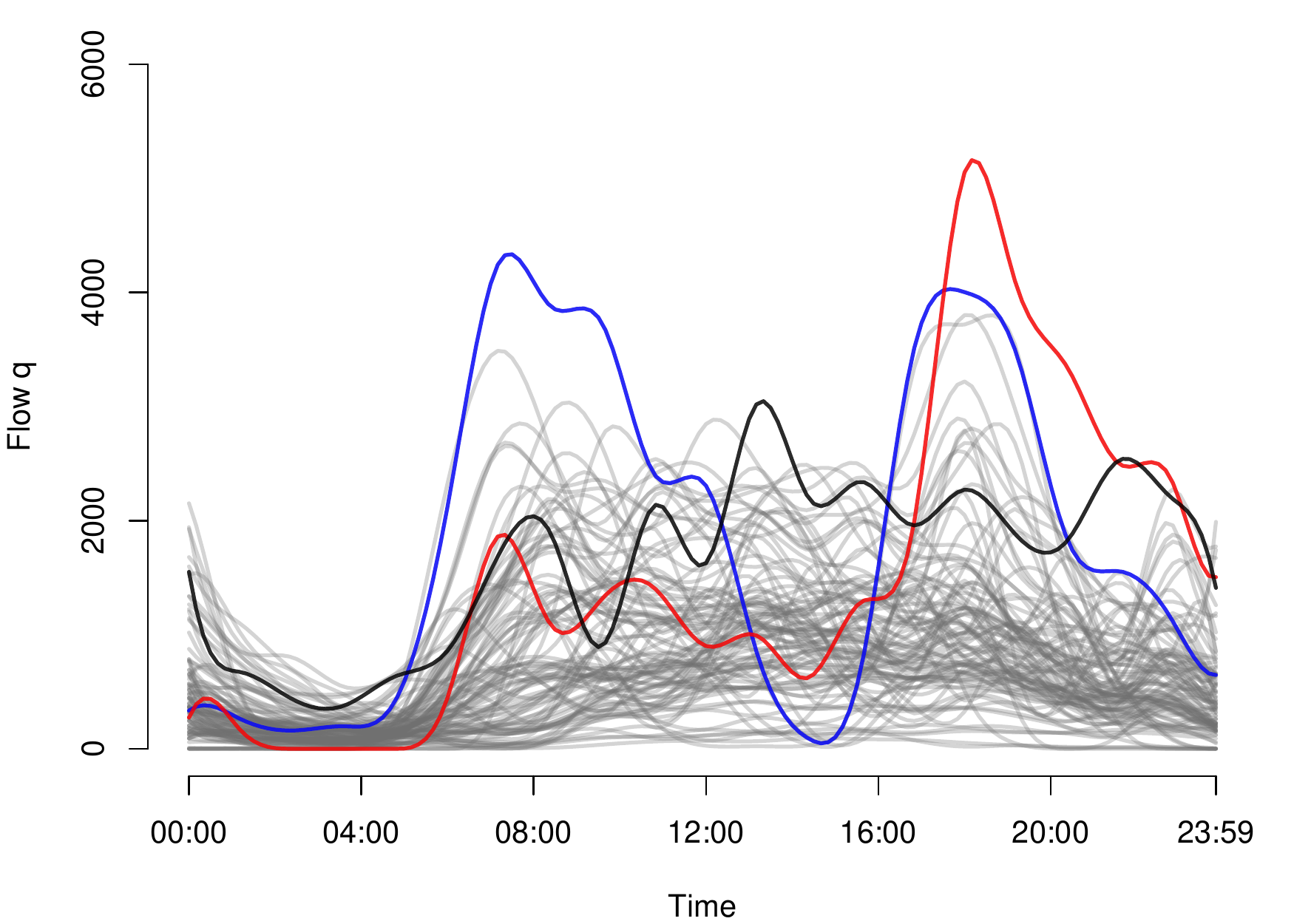}
	\end{center}
	\caption{\scriptsize{Daily traffic flow profiles as curves. Each observation is an M-spline-smoothed version of the day-wise  traffic counts measured in 1-minute-intervals and transformed to hourly basis as specified in (\ref{hrs}).}}
\end{figure}

The notions of sample quantiles and expectiles are useful when analyzing extremes of univariate data. When dealing with high dimensional data, it is not clear how to define what is extreme. When we think of multivariate data as curves sampled on some regular grid, we can utilize many definitions of 'extreme' behavior. We could, for example, consider wiggly curves as extremes or the curves that lie above or below most of the other curves. 

Daily traffic profiles composed from $p$ traffic flow measurements per day on $n$ days present such multivariate data. Consider 100 randomly chosen smoothed daily flow profiles, comprising $p=144$ vehicle counts (converted to hourly equivalent) over a day as collected from the roundabout on Ernst-Reuter-Platz in the city center of Berlin, Germany in Fig. (\ref{intro_lines}). One example of an extreme pattern would be the path highlighted in blue on Fig.  (\ref{intro_lines}), which corresponds to the highest average load, another example would be the red path, which represents the profile with the highest evening-peak, yet another example is the black profile with the highest load at noon.

What we considered extreme in Fig. (\ref{intro_lines}) were the extremes of some real numbers computed as (very simple)  linear combinations of the traffic flow.

Which in turn means that the tail indices of linear combinations of multivariate data may hold additional information about the data set. But how to choose the linear combination that suits best? A natural choice would be a projection of the data along a direction which fulfills some meaningful criteria. 

Indeed, an elegant and coherent definition of 'extremeness' for high dimensional data is as a projection-based one.  \cite{hallin}, \cite{kongMizera},\cite{FRAIMAN2012} and \cite{MONTESROJAS201720} among others study  different specifications of directional (or projection-based) quantiles.

\cite{TRAN20191} choose the concept of expectiles to obtain their directional interpretation of 'extremeness' in form of directional expectiles. They naturally extend the classical PCA by replacing the mean by  its expectile generalization,  and the variance to maximize -  by its asymmetrically weighted expectile based analogue. The proposed algorithm finds an orthonormal basis, in which each principal expectile component (PEC) determines the directional expectile to capture interesting structures in the data extremes. 

In this paper, we apply the method of \cite{TRAN20191} which they term "Principal component analysis in an asymmetric norm"  to reduce the dimensionality of traffic flow profiles with focus on their directional extremes. We extend thereby the related methodological basis for dimension reduction of traffic flow profiles,  proven useful in \cite{COOGAN}, \cite{CRAWFORD2017}, and \cite{GUARDIOLA2014119} among others.
Using the traffic flow data collected from January to July 2018 with 14 detectors in different points of the Ernst-Reuter-Platz roundabout in the city center of Berlin, Germany, 
we estimate the principal expectile components (PECs) from the data, visualize and interpret the results. We 
provide guidelines for identification and interpretation of the resulting directional extremes and argue, that our approach clearly identifies some outstanding patterns in the traffic flow.

The structure of this paper is as follows. In the next section, we give the methodological details of the proposed technique, describe the data and our preprocessing steps. Then, we present the results, visualize and interpret them. We give the intuition behind the extracted components, show how to identify the extreme patterns and how to relate them to additional information as the location and the day type. Finally, we conclude and discuss our findings. 

\section{Methods and Data}
In this section we introduce the method we are using and describe the data. We start with reviewing the notions of quantiles and expectiles, then we go on with the classical PCA and its asymmetric version of \cite{TRAN20191}. We also present the algorithms for computing the expectiles and the associated principal expectile components.
The last subsection is devoted to the data description and our preprocessing steps.

\subsection{Quantiles and expectiles}
The classical extremeness measures for a univariate data are quantiles (\cite{koenker}). For a real valued random variable $X$ with the cumulative distribution function $F_X$ and $0<\tau<1$, the $\tau$-quantile, $q_{\tau} = q_\tau(X)$ is the inverse of the cumulative distribution function of $X$ in $\tau$, formally  $q_{\tau}= F_X^{-1}(\tau)$.

The $\tau$-quantile can be also defined as the minimizer of the  expectation of the following asymmetric norm:
\begin{align}\label{qt}
q_{\tau} = \argmin\limits_{q\in\mathbb{R}}\rE\|X-q\|^1_{\tau,1} 
\end{align}
where for $\alpha=1,2$ an asymmetric $\ell_\alpha$-norm:
\begin{align}\label{asymnorm}\|x\|^\alpha_{\tau,\alpha} = |x|^\alpha\cdot \{\tau \mathbf{1}_{x \geq 0} + (1-\tau) \mathbf{1}_{x < 0}\}
\end{align}
In (\ref{qt}) an asymmetric $\ell_1$-norm is used.

Quantiles have been used in the context of traffic analysis mainly for accessing travel time reliability (e.g. in \cite{ZANG}, \cite{LI2019}, and \cite{LI2020}), and for traffic flow predictions in \cite{COOGAN2}.

\begin{figure}[ht]
\centering
\includegraphics[width=0.5\textheight]{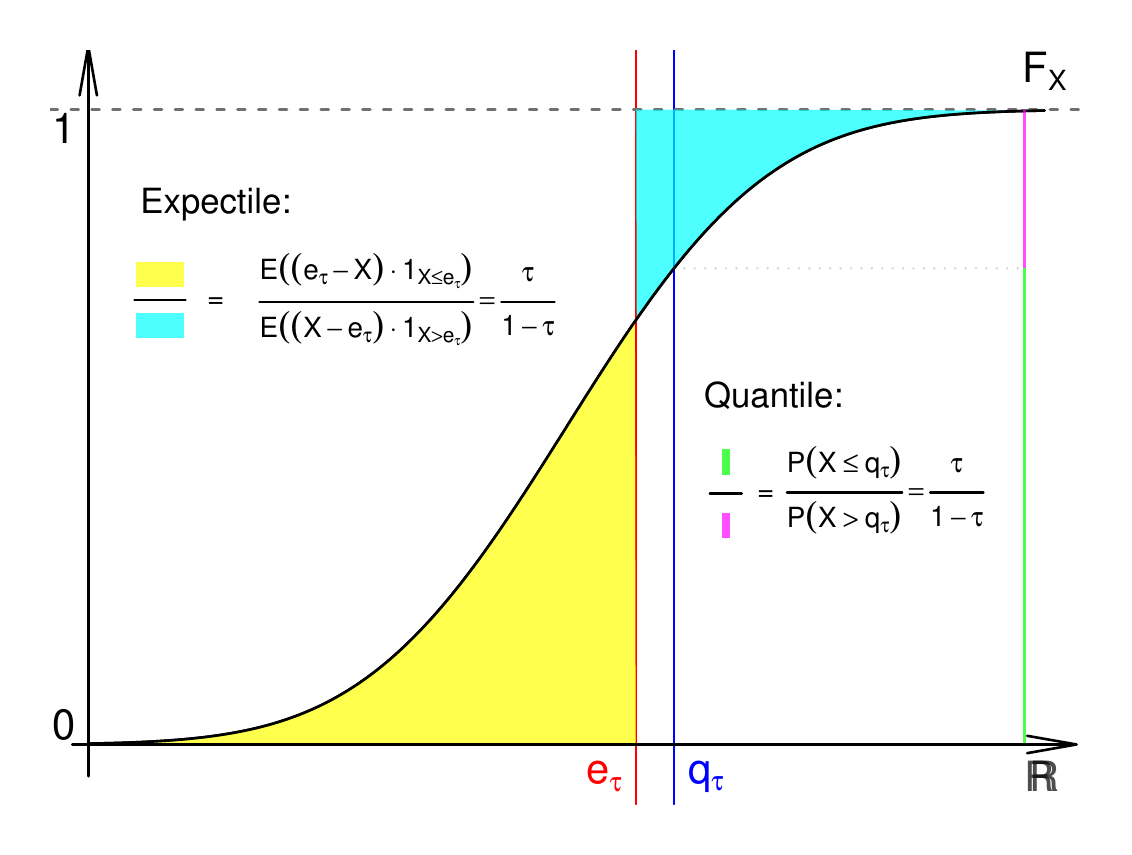}
\caption{\scriptsize{The $\tau$-quantile $q_\tau$ and the $\tau$-expectile $e_\tau$ of a real-valued random variable $X$ with strictly increasing cumulative distribution function $\sf{F}_X$ ($\tau=0.75$). $\sf{P}$ denotes probability, $1_A$ denotes indicator function on $A$, $\rE$ denotes expectation with respect to the distribution of $X$.}}\label{qe}
\end{figure}

A tail measure similar to the $\tau$-quantile is the $\tau$-expectile introduced in \cite{newey}. The $\tau$-expectile is the real number, such that the $\tau$-proportion of the expected distance to it lies below and the $(1-\tau)$-proportion lies above it. As quantiles generalize the concept of the median, expectiles generalize the concept of the mean.  A comparative visualization of both concepts for a real-valued random variable $X$ with strictly increasing cumulative distribution function $\sf{F}_X$, related to the corresponding first order conditions for (\ref{qt}) and (\ref{et}), and 
 inspired by Fig. 9 in \cite{Rowland2019} is shown in Fig. (\ref{qe}).

The $\tau$-expectile, denoted as $e_\tau = e_\tau(X)$, can be defined as the minimizer of the expected distance to $X$ in an asymmetric $\ell_2$-norm:
\begin{align}\label{et}
e_\tau = \argmin\limits_{e\in\mathbb{R}}\rE\|X-e\|^2_{\tau,2}.
\end{align}
While the $\tau$-quantile is based on the asymmetric $\ell_1$-norm in (\ref{qt}), the $\tau$-expectile expression contains the asymmetric $\ell_2$-norm. For $\tau=0.5$ we recover the symmetry and $e_{0.5}$ corresponds to the mean.

Expectiles have been used in numerous economic applications (\cite{taylor}, \cite{kokic}, \cite{Schnabel_21_5}). We are not aware of any reference on using expectiles for traffic data analysis.

\subsection{PCA in an asymmetric norm}

The concepts of quantiles and expectiles are suitable for measuring extremeness of univariate data. When dealing with high dimensional data, it is not clear how to define what is extreme. An elegant and coherent definition of 'extremeness' for such data is as a projection-based one (see, for example \cite{hallin}, \cite{kongMizera},\cite{FRAIMAN2012} and \cite{MONTESROJAS201720} for projection-based quantiles).

\cite{TRAN20191} utilize the concept of expectiles to obtain their directional interpretation of 'extremeness' in form of directional expectiles. They naturally extend the classical PCA by replacing the mean by  its expectile generalization,  and the variance to maximize -  by its asymmetrically weighted expectile based analogue. 
In this paper, we use the approach of \cite{TRAN20191}  for dimensionality reduction of daily traffic flow profiles with a focus on their extremes as it is a natural extension of existing PCA-based dimension reduction techniques,  proven useful in \cite{COOGAN}, \cite{CRAWFORD2017}, and \cite{GUARDIOLA2014119} among others.

The classical PCA  (\cite{pca})  is a  very popular and powerful technique for projecting multivariate data in a low dimensional subspace with easily obtainable and interpretable results.  The principal components are usually computed as the eigenvectors of the covariance matrix of the data and point in the direction of the largest variation around its mean. Projected on a principal component direction, each observation in $\mathbb R^p$ becomes a real valued principal component score; the variation of the scores is the greatest possible. Suppose, we observe the samples $y_1, y_2,\ldots y_n\in \mathbb R^p$ of $\mathbb R^p$-valued random variable $Y$. Then, the first principal component $\phi^*$ is
\begin{align}
\phi^* = \argmax_{\phi \in \mathbb R^p, \phi^\top\phi = 1} \textup{Var}(\phi^\top Y).
\end{align}
The sample version of the classical PC maximizes the sample variation of the projected points $\{\phi^\top y_i\}_{i=1}^{n}$ around their sample mean $\bar y$:
\begin{align}
\hat\phi^* = %\argmax_{\phi \in \mathbb R^p, \phi^\top\phi = 1}\widehat{\textup{Var}}(\phi^\top y_i,i=1,\ldots,n)
\argmax_{\phi \in \mathbb R^p, \phi^\top\phi = 1}\frac{1}{n}\sum_{i=1}^n(\phi^\top y_i - \bar y)^2.
\end{align} 

\cite{TRAN20191} define their principal expectile component (PEC) as the unit vector in the direction that maximizes the $\tau$-variance of the projection.  
 For a real valued random variable $X$, the $\tau$-variance $\textup{Var}_\tau(X)$ is the expectation of the squared deviation of $X$ to its $\tau$-expectile in an asymmetric $\ell_2$-norm, formally:
\begin{align}
\textup{Var}_\tau(X) = \rE||X - e_\tau(X)||^2_{\tau,2}
\end{align}
Consequently, the PEC for a certain $\tau$, denoted as $\phi_\tau^\ast=\phi_\tau^\ast(Y)$, is the unit vector that maximizes $\textup{Var}_\tau(\phi^\top Y)$:
\begin{align}
\phi_\tau^\ast = \argmax_{\phi \in \mathbb R^p, \phi^\top\phi = 1}\textup{Var}_\tau(\phi^\top Y) 
= %\argmax_{\phi \in \mathbb R^p, \phi^\top\phi = 1}\textup{E}\{\tau(\phi^\top Y - e_\tau)^2 \mathbbm{1}_{(\phi^\top Y>e_\tau)} +(1-\tau)(\phi^\top Y - e_\tau)^2 \mathbbm{1}_{(\phi^\top Y<e_\tau)}\},
\argmax_{\phi \in \mathbb R^p, \phi^\top\phi = 1}\rE||\phi^\top Y - e_\tau||^2_{\tau,2}
\end{align}
where $e_\tau \in \mathbb R$ is the $\tau$-expectile of $\phi^\top Y$ and $\|\cdot\|^\alpha_{\tau,\alpha}$ the asymmetric $\ell_2$-norm from (\ref{asymnorm}).

The sample counterpart to $\phi_\tau^\ast$, denoted as $\hat\phi_\tau^\ast$, can be obtained from the observed values $y_1,\ldots, y_n$ as:
\begin{align}\label{phiast}
\hat\phi_\tau^\ast %&= \argmax_{\phi \in \mathbb R^p, \phi^\top\phi = 1}\textup{Var}_\tau(\phi^\top y_i,i=1,\ldots,n) \\
&= \argmax_{\phi \in \mathbb R^p, \phi^\top\phi = 1}\frac{1}{n}\sum_{i=1}^n(\phi^\top y_i - \hat e_\tau)^2w_i,
\end{align}
where $\hat e_\tau$ is the sample $\tau$-expectile, computed from the real valued scores
$\phi^\top y_1, \ldots \phi^\top y_n$ via Algorithm \ref{alg:e}, and 
\begin{align}
w_i = \left\{ \begin{array}{cc} \tau & \mbox{ if } \sum_{j=1}^p y_{ij}\phi_j > \hat e_\tau, \\ 1-\tau & \mbox{ otherwise.} \end{array} \right.
\end{align}
are the corresponding asymmetric weights. Given these weights $w_i$, we can group the data points into two sets $\mathcal I_\tau^+$ and $\mathcal I_\tau^-$ depending on whether they are given the weight $\tau$ or $(1-\tau)$ respectively:
\begin{align}
\mathcal I_\tau^+ = \{i \in \{1, \ldots, n\}: w_i = \tau\}, \mathcal I_\tau^- = \{i \in \{1, \ldots, n\}: w_i = 1-\tau\}.
\end{align}
\begin{algorithm}[h]\label{alg:e}
	\vspace{0.2cm}\hrule\vspace{0.2cm}
	\KwData{$X\in\mathbb R^{n\times 1}$}
	\KwResult{an estimated expectile $\hat e_\tau$ of $X$.}
	Initialize the weights $w_i^{(0)}, i=1,\ldots, n$\\
	Set $t=0$\\
	
	\Repeat{$w_i^{(t)} = w_i^{(t+1)}$for all $i$}{Let $\mathcal I_\tau^{+(t)}$ be the set of indices $i$ such that $w_i^{(t)} = \tau$, and $\mathcal I_\tau^{-(t)}$be the complement\;
		Set $n^{+(t)}=|\mathcal I_\tau^{+(t)}|$ and $n^{-(t)}=|\mathcal I_\tau^{-(t)}|$\;
		Compute $e_\tau^{(t)} = \frac{\tau\sum_{i \in \mathcal I_\tau^{+(t)}}X_i + (1-\tau)\sum_{i \in \mathcal I_\tau^{-(t)}}X_i}{ \tau n^{+(t)} + (1-\tau)n^{-(t)}}$ \;
		Update $w_i$: \\
		\eIf{$X_i>e_\tau^{(t)}$}{
			set $w_i^{ (t+1)} = \tau$\\
		}{
			set $w_i^{ (t+1)} = 1-\tau$\;
		}
		Set $t = t + 1$.}
	
	\Return{$\hat e_\tau = e_\tau^{(t)}$}
	\vspace{0.2cm}\hrule\vspace{0.2cm}
	\caption{Asymmetric weighted least squares for computing $\hat e_\tau$ from \cite{newey}.}
\end{algorithm}
If $\tau>0.5$, then the observations in $\mathcal I_\tau^+$ are weighted  in (\ref{phiast}) higher (by $\tau$), whereas the observations in $\mathcal I_\tau^-$   are weighted lower (by $(1-\tau)$). If $\tau<0.5$, the vice versa is the case: the observations in $\mathcal I_\tau^+$ are weighted lower and the observations in $\mathcal I_\tau^-$  a higher impact on (\ref{phiast}).

\cite{TRAN20191} show that their PEC is the classical PC of a weighted version of the covariance matrix. This property can be used to efficiently compute $\hat\phi_\tau^\ast$. 

Let $n_+=|\mathcal I_\tau^+|$ and $n_-=|\mathcal I_\tau^-|$ be the sizes of the sets containing data points with different weights. We define the centering constant $\tilde e_\tau$, which may be different from the sample mean for $\tau\not=0.5$, as:
\begin{equation}\label{tet}
\tilde e_\tau = \frac{\tau\sum_{i \in \mathcal I_\tau^+}y_i + (1-\tau)\sum_{i \in \mathcal I_\tau^-}y_i}{ \tau n_+ + (1-\tau)n_-},
\end{equation}
and establish the asymmetrically weighted sample covariance matrix $C_\tau$ centered at $\tilde e_\tau$ as:
\begin{equation}\label{eqn:C}
C_\tau = \frac{\tau}{n}\left\{\sum_{i\in \mathcal I_\tau^+}(y_i-\tilde e_\tau)(y_i-\tilde e_\tau)^\top\right\} + \frac{1-\tau}{n}\left\{\sum_{i\in \mathcal I_\tau^-}(y_i-\tilde e_\tau)(y_i-\tilde e_\tau)^\top \right\}.
\end{equation}
Then, the first sample PEC $\hat\phi^\ast_\tau$ is then the largest eigenvector of $C_\tau$.
The resulting iterative procedure to find $\hat\phi_{\tau}^\ast$ is recorded in Algorithm \ref{alg:fi}.

\begin{algorithm}[ht]\label{alg:fi}
	\vspace{0.5cm}\hrule\vspace{0.5cm}
	\KwData{$Y\in\mathbb R^{n\times p}$}
	\KwResult{a vector $\hat{\phi}$, an estimator of the first principal expectile component of $Y$.}
	Initialize the weights $w_i^{(0)}$\\
	Set $t=0$\\
	\Repeat{$w_i^{(t)} \not= w_i^{(t+1)}$for all $i$}{Let $\mathcal I_\tau^{+(t)}$ be the set of indices $i$ such that $w_i^{(t)} = \tau$, and $\mathcal I_\tau^{-(t)}$be the complement\;
		Set $n^{+(t)}=|\mathcal I_\tau^{+(t)}|$ and $n^{-(t)}=|\mathcal I_\tau^{-(t)}|$\;
		compute $\tilde e_\tau^{(t)}$ as in equation \ref{tet} with sets $\mathcal I_\tau^{+(t)}$ and $\mathcal I_\tau^{-(t)}$\;
		compute $C_\tau^{(t)}$ as in equation \ref{eqn:C} with sets $\mathcal I_\tau^{+(t)}$ and $\mathcal I_\tau^{-(t)}$\;
		set $\phi^{(t)}_\tau$ to be the largest eigenvector of $C_\tau^{(t)}$\;
		set $e_\tau^{(t)}$ to be the $\tau$-expectile of $(\phi^{(t)})^\top Y$\;
		update $w_i$: \\
		\eIf{$(\phi^{(t)}_\tau)^\top Y_i>e_\tau^{(t)}$}{
			set $w_i^{ (t+1)} = \tau$\\
		}{
			set $w_i^{ (t+1)} = 1-\tau$\;
		}
		set $t = t + 1$.
	}
	\Return{$\hat\phi_{\tau} = \phi^{(t)}_\tau$}
\vspace{0.2cm}\hrule\vspace{0.2cm}
	\caption{Principal expectile algorithm from \cite{TRAN20191}.}
\end{algorithm}

The presented algorithm \ref{alg:fi} computes only the first principal expectile component which is further denoted as $\hat\phi_{\tau1}$ for convenience. To obtain the second PEC, $\hat\phi_{\tau 2}$, we can simply repeatedly apply  the algorithm to the residuals
$y_i - \hat\phi_{\tau1}(\hat\phi_{\tau1}^\top y_i + \hat e_{\tau1})$, where $\hat e_{\tau1}$ is the $\tau$-expectile of the scores $\hat\phi_{\tau1}^\top y_i$. The procedure remains the same for higher order components.

\subsection{Data and preprocessing} 
Our data consists of traffic counts measured daily in 1-minute intervals from the January, 10, 2018 to July 11, 2018 (February is missing for unknown reasons) via 14 visual detectors installed on the heavily loaded roundabout on Ernst-Reuter-Platz in the city center of Berlin, Germany. \footnote{\href{http://flow.dai-labor.de/datasets/}{http://flow.dai-labor.de/datasets/}, accessed on 10. Mai 2020}  We consider the traffic flow profiles composed of the traffic flow counts over each available day.

Let us denote the minute-wise vehicle counts as $c_{d\ell t}$ with $t=1,\ldots 1440$ indexing all 1-minute intervals throughout a day, $d=1, \ldots, 192$ indexing available days, and $\ell=1,\ldots 14$ indexing the detector locations. 
We transform $c_{d\ell t}$ to its hourly equivalent, $\tilde q_{d\ell t}$:
\begin{align}\tilde q_{d\ell t} = \frac{c_{d\ell t}\cdot 3600}{T},\label{hrs}
\end{align}
where $T=60$ is the duration of the aggregation interval  in seconds.
All daily profiles with consecutive zero counts for periods of more than six hours are removed, because such long sequences of zeros are atypical for the considered  heavy loaded roundabout and may indicate a temporal road closure or a malfunction of the detector.
The dimension of the remaining data counts $n=d\cdot \ell=1656$
'location-day' flow profiles $\{\tilde q_{it}\}_{t=1}^p$, with $i=1,\ldots, n$ and $p=1440$ measurements over a day.
\begin{figure}[H]
	\label{map1}
	\centering
	\includegraphics[width=0.65\textwidth]{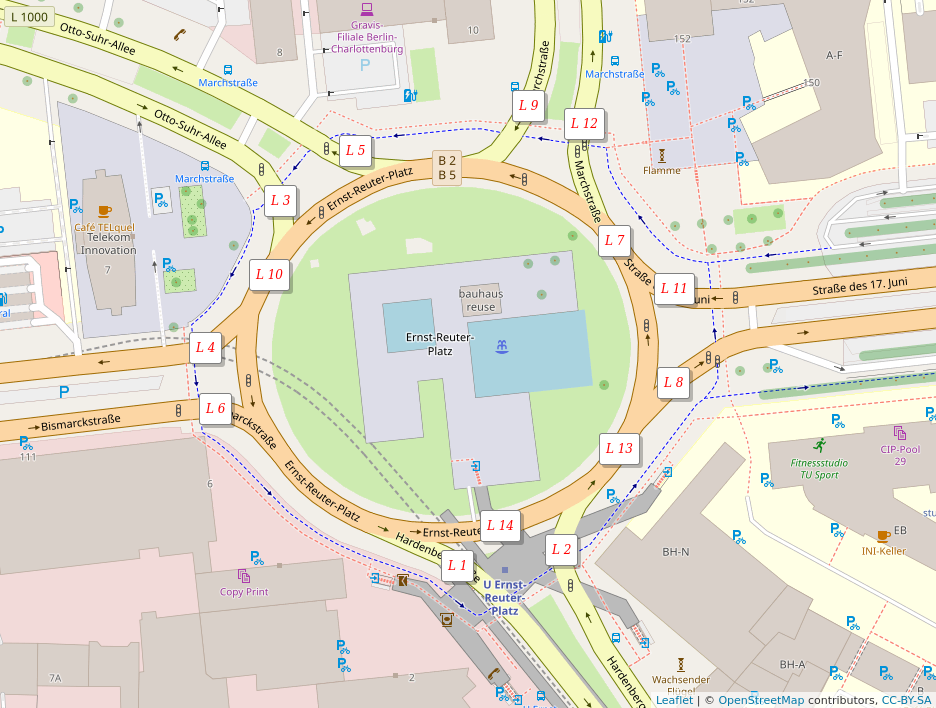}
	\caption{\scriptsize{The locations of the 14 visual detectors on the roundabout infrastructure point on Ernst-Reuter-Platz in Berlin, Germany  generated the considered traffic counts data.}} 
\end{figure}

\begin{table}[ht]
	\centering
	\begin{tabular}{rllllll}
	\hline
	& 0:00 - 03:59 & 04:00 - 07:59 & 08:00 - 11:59 & 12:00 - 15:59 & 16:00 - 19:59 & 20:00 - 23:59 \\ 
	\hline
	Mon & 157.3(266.9) & 285.4(532.9) & 724.2(873.1) & 854(907.2) & 981.4(1008.2) & 590.7(708.1) \\ 
Tue & 180.1(283.6) & 294.6(586.4) & 988.8(970.4) & 918.8(868.2) & 971.3(1034.9) & 652.3(757) \\ 
Wed & 194.8(287.8) & 293.4(501.7) & 1035.3(940.2) & 1087.7(908.4) & 1168.5(1095.7) & 645.1(776.8) \\ 
Thu & 178.6(279.3) & 272.9(488.8) & 968.6(991.5) & 977.2(948.9) & 981.4(1089.7) & 586.4(778.4) \\ 
Fri & 199.5(318.6) & 248.7(469.3) & 769.8(937.1) & 962.9(954.4) & 958.3(1056.3) & 617.1(809.7) \\ 
Sat & 270.5(423.1) & 146.4(253.3) & 367.7(451.4) & 721.8(726.7) & 804.5(904.3) & 642.7(802.8) \\ 
Sun & 365.4(538.4) & 171.9(286.9) & 254.4(365.2) & 539.3(624.4) & 614.8(725.4) & 464.5(601.2) \\ 
	\hline
	\end{tabular}
	\caption{{The means (the standard deviations) of the traffic counts for different days of week and day times.}} 
	\label{sum}
\end{table}

Figure \ref{map1} shows the locations of the 14 visual detectors on the roundabout infrastructure point on Ernst-Reuter-Platz in Berlin, Germany and Table \ref{sum} presents the day-type and time-of-day-type summary statistics for the data across $n$ flow profiles. A closer look at the data summary in Table \ref{sum} reveals, that there are persistent patterns in traffic flow variability due to the day-type and the time-of-day particularities. 

We smooth the noisy daily flow profiles  $\{\tilde q_{it}\}_{t=1}^p$,  $i=1,\ldots, n$, using M-splines (\cite{ramsay1988}) with non-negative coefficients in order to insure the non-negativity of the resulting smoothed flow profiles\footnote{We use  M-splines of degree 3 as implemented by the function mSpline() in the R package splines2 (https://CRAN.R-project.org/package=splines2 ) and enforce the non-negativity of the coefficients using the function penalized() in the R package penalized (https://CRAN.R-project.org/package=penalized)}.
We then evaluate the resulting daily traffic flow curves on a discrete time grid of 10-minute intervals to obtain the matrix $Q$ with the  dimensions $p\times n$, where $p=144$ and $n= d\cdot \ell = 1656$
  for our subsequent analysis. 
There are $n$ columns in $Q$ corresponding to the $n$ smooth daily traffic flow profiles each of length $p$, denoted as $q_i = \{q_{it}\}_{t=1}^p$, $i=1,\ldots, n$. $Q$ contains the preprocessed traffic flow profiles originating from all 14 cameras installed on the roundabout in the considered time period.
In the following, we consider the column vectors of $Q$, $q_i,i=1,\ldots, n$,  representing the daily traffic flow profiles as discretized curves (as in Fig. (\ref{intro_lines})). In our subsequent analysis, we are going to look for the projection directions that best discover directional extremes of the traffic profiles in $Q$. 

\section{Directional extremes of daily traffic flow profiles}
In this section we apply the methodology of \cite{TRAN20191} to the preprocessed traffic flow profiles in $Q$ to identify their directional extremes.
That is, we are going to find the corresponding vectors of length $p$, our principal expectile components, such that, projected on these vectors the traffic profiles are represented in a low dimensional space of the first $k$ PECs in a way that the appropriately weighted variation of the projection around its expectile is the largest possible. Thereby we weight the traffic profiles in $Q$ asymmetrically, meaning that some traffic profiles are weighted with $\tau$ and some  with $1-\tau$. This way we naturally obtain directional extreme trajectories of the traffic profiles.

Precisely, by executing the steps of the Algorithm \ref{alg:fi} for a specific $\tau\in(0,1)$, we compute the first sample PEC, denoted as $\hat\phi_{\tau1}$,  that corresponds to the projection direction, in which the projection (which is a linear combination of each $\{q_{it}\}_{t=1}^p$ with weights determined by $\hat\phi_{\tau1}$)
has the highest $\tau$-variance. For each flow profile $q_i$, we then compute its score $\hat\phi_{\tau1}^\top q_i$: the projection of $q_i$ on the first PEC for the chosen $\tau$. Each score is the optimal (with respect to the $\tau$-variance maximization) one dimensional representation of the corresponding flow profile. By maximizing the asymmetric $\tau$-variance we put our focus on the tails of the projection and the direction obtained reveals the most informative structure in the data extremes in the $\tau$-variance sense. Taking the residuals of the first PEC, $q_i - \hat\phi_{\tau1}(\hat\phi_{\tau1}^\top q_i + \hat e_{\tau1})$, $i=1,\ldots,n$ with   $\hat e_{\tau1}$ being the $\tau$-expectile of the scores $\{\hat\phi_{\tau1}^\top q_i\}_{i=1}^n$, we compute the second PEC, denoted as $\hat\phi_{\tau2}$, and the associated scores $\hat\phi_{\tau2}^\top q_i$ for each 'location-day' $i=1,\ldots,n$ in this second direction. The iterative procedure is repeated untill the intended number ($K$) of PECs is estimated and $\hat\phi_{\tau,K}$ is obtained. 

\begin{figure}[hb!]
	\centering
	\includegraphics[scale=0.4]{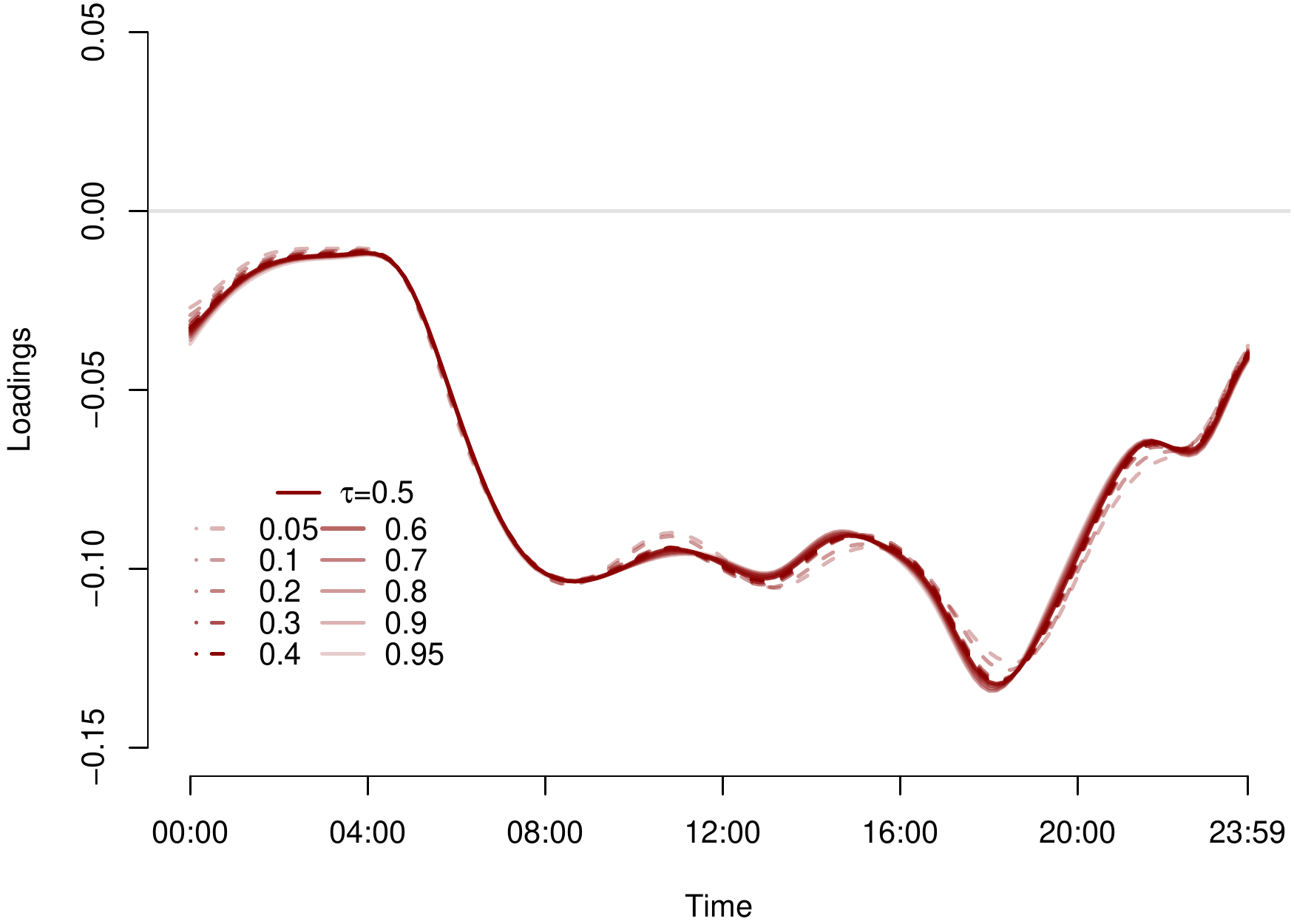}\includegraphics[scale=0.4]{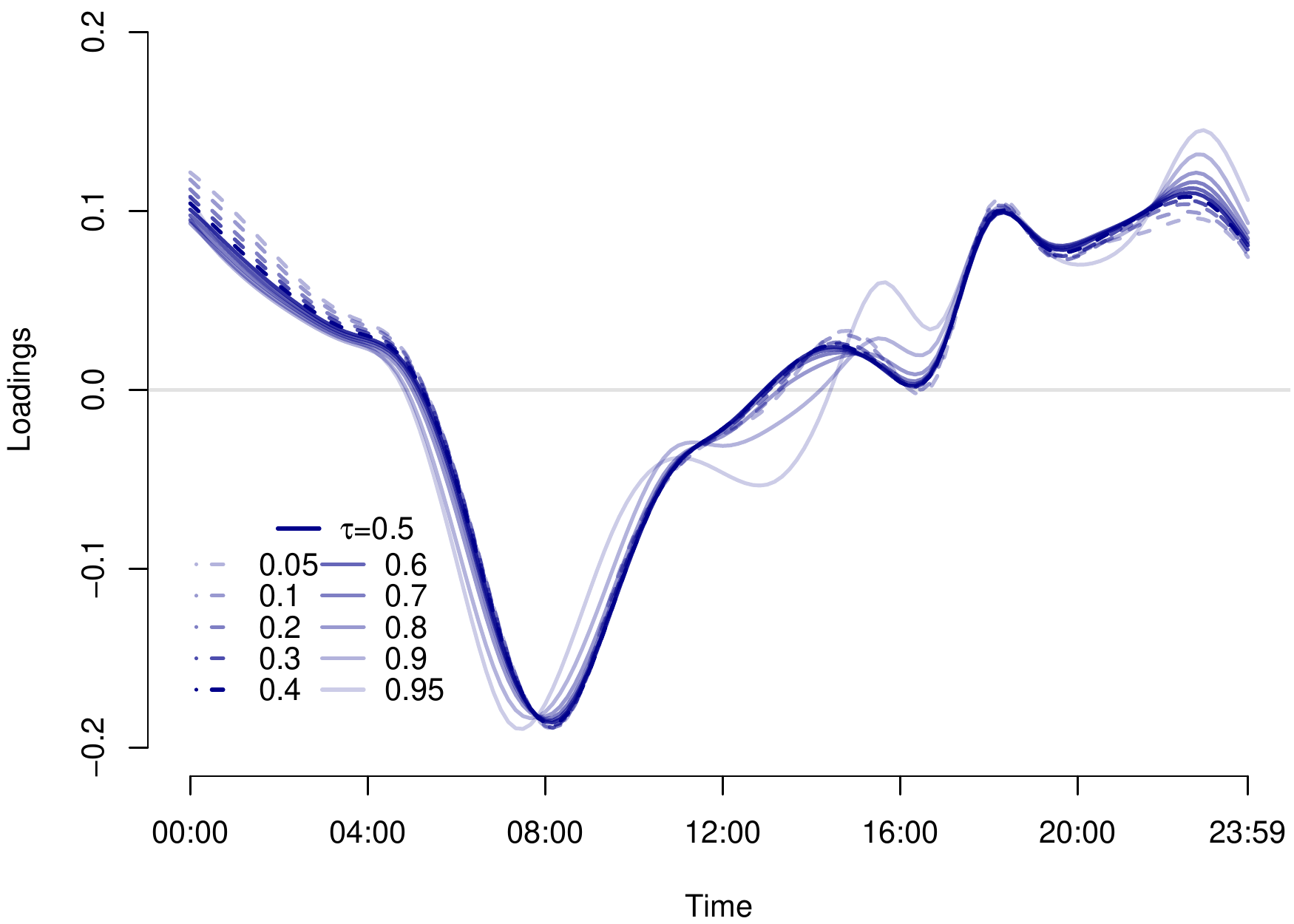}\\
\includegraphics[scale=0.4]{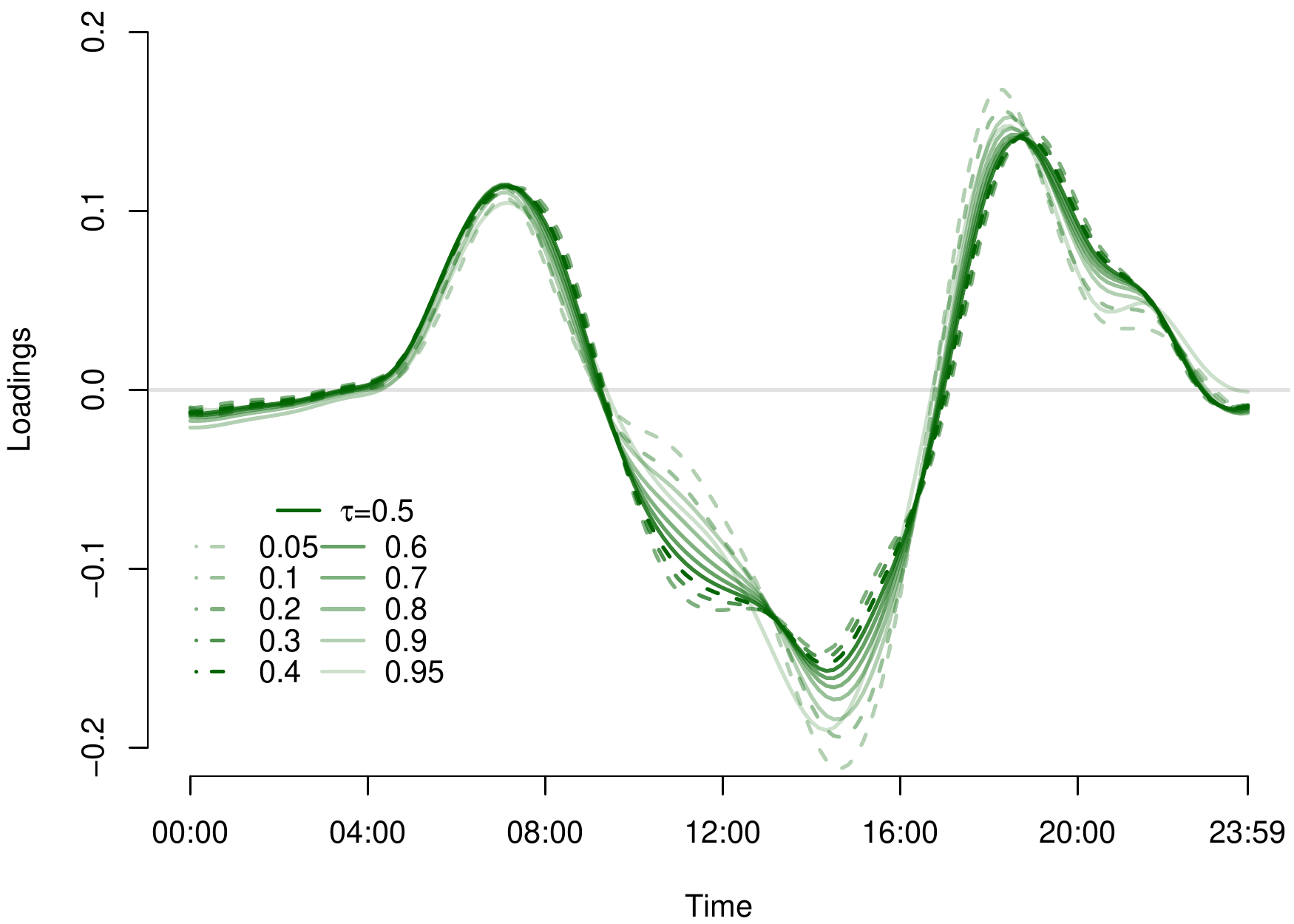}\includegraphics[scale=0.4]{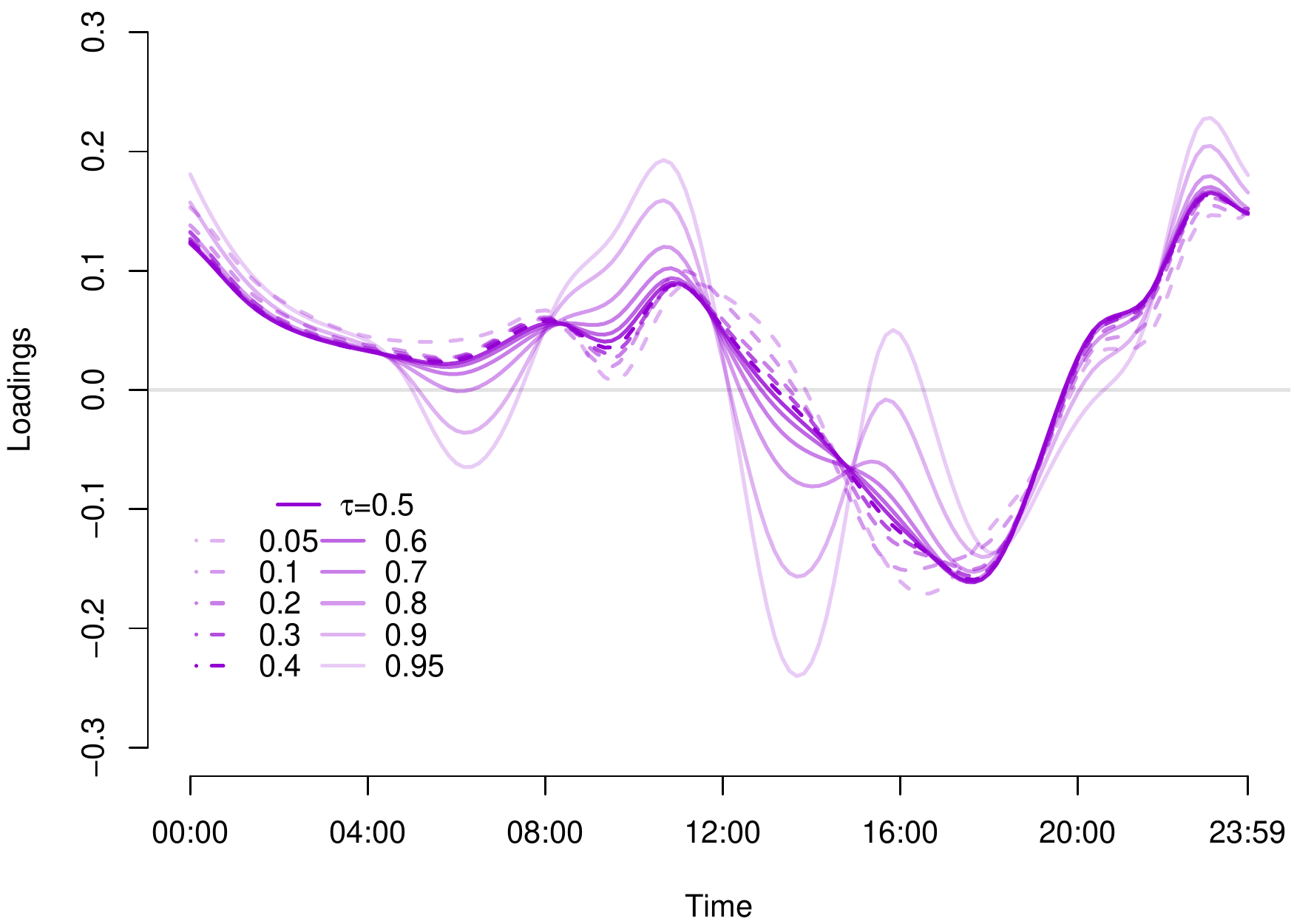}
	\caption{\scriptsize{1st (red shades, top left), 2nd (blue shades, top right),  3nd (green shades, bottom left), and  4th (purple shades, bottom right)  PECs of the traffic flow curves for $\tau$ from 0.05 to 0.45 (light to dark shades, dashed) and from 0.5 to 0.95 (dark to light shades, solid).}}	
\label{pec}	
\end{figure} 

Unfortunately, there is no feasible criterion for determining the number of components to keep. We align our choice with the previous studies using dimensionality reduction for traffic profiles.
\cite{GUARDIOLA2014119} analyse data from a single detector and keep the first three components. We consider a collection of traffic profiles from multiple detector locations, similar to \cite{COOGAN} who choose four. Thus we also keep the first four and the "plus one" component to the three chosen in \cite{GUARDIOLA2014119} accounts (as seen later in the section) for the differences in the traffic flow levels between the locations. Consequently, $K=4$ and we have to repeat the steps in algorithm \ref{alg:fi} four times on the residuals to obtain the first $K$ PECs: $\hat \phi_{\tau, k}, k=1,\ldots, K$.

To be able to compute the PECs we also have to specify the expectile levels, which we choose to set to multiple values in $(0,1)$: 
 $\tau=$ 0.05, 0.1, 0.2,$\ldots$, 0.9, and 0.95. The case of $\tau=0.5$ corresponds to the usual principal component (PC). The farther is the chosen expectile level $\tau$ from 0.5, the deeper is the focus shifted to the tails of the projection and the more 'extreme' (with respect to $\hat \phi_{\tau k}$) are the daily profiles $\{q_i\}_{i: w_i=\tau}$ weighted higher in (\ref{tet}). If $\tau>0.5$, the profiles contained in $\mathcal I^+_{\tau,k}$ are weighted higher (by $\tau$) as opposed to all the other paths weighted by $1-\tau$ and contained in $\mathcal I^-_{\tau,k}$. If $\tau<0.5$ the vice versa is the case:  the profiles contained in $\mathcal I^-_{\tau,k}$ become the higher weight ($1-\tau$) as opposed to all the other weighted by $\tau$ and contained in $\mathcal I^+_{\tau,k}$. 

The first $K$ estimated  PECs for the traffic profiles in $Q$ are shown in Fig. \ref{pec}.
The interpretation of the PECs in Fig. \ref{pec} is as follows.
The first PEC points in the direction that accounts for the overall (detector location) specific traffic flow level. A positive score 
on this component means, that the corresponding traffic profile has lower flow values compared to the central tendency in the whole collection $Q$. On the contrary, a negative score distinguishes the profiles with relatively high flow level. Note, that changing the $\tau$-level does not notably influence the shape of the first component, showing that in this case, the direction of the largest $\tau$-variance coincides with the direction of the largest variance in the data. However, the corresponding central tendency (which builds the basis for comparison of trajectories) $\tilde e_\tau$, specified  in (\ref{tet}) and  used for centering the weighted covariance matrix in  (\ref{eqn:C}) prior to computing the PECs, still changes with varying $\tau$. For $\tau=0.5$,  $\tilde e_{\tau=0.5}$ is the average trajectory of the traffic profiles. 
For $\tau\not=0.5$, e.g. $\tau = 0.95$, $\tilde e_{\tau}$ is the weighted average of the traffic flow profiles $q_i, i=1,\ldots,n$ weighted either with $\tau=0.95$ if  or with $1-\tau = 0.05$, formally:
\begin{equation}\label{tet0.95}
\tilde e_{\tau=0.95}= \frac{0.95\sum_{i \in \mathcal I_\tau^+}q_i + 0.05\sum_{i \in \mathcal I_\tau^-}q_i}{ 0.95 n_+ + 0.05n_-}.
\end{equation}
This constant updates as soon as we compute subsequent PECs, since the sets $\mathcal I_\tau^+$ and $\mathcal I_\tau^-$ (essentially $\mathcal I_{\tau, k}^+ $ and $ \mathcal I_{\tau, k}^- $ as we omitted index $k$ before) change with  altering  $k$. For completeness, we add the index $k$ also to the centering constant from now on.
The resulting centering constants $\tilde e_{\tau, k}$ express the $\tau$ weighted $k$th component augmented central tendency of the traffic profiles in $Q$.

The second PEC measures how pronounced is the peak in the early rush hour around 08:00 relative to the rest of the day. There are some minor differences in the component shapes for different $\tau$s observed mainly in the night hours and in the small 'bump' around 16:00. A negative score on this component indicates a traffic profile with rather intensive flow around the early rush hour and the opposite elsewhere, both relative to the central tendency in the collection considered.

The third PEC relates the peaks around the early and the late rush hours to the mid-day load. A traffic profile scored positive on this component demonstrates high flow during the rush hours and low flow during the mid-day (compared to the central tendency of all profiles). We observe some disaccordance of the component shapes when $\tau$ alters. Some notable differences show up in the late rush hour between 16:00 and 20:00, in the mid-day flow between 11:00 and 12:00, and in the afternoon between 14:00 and 16:00.

The fourth PEC exhibits the most pronounced shape changes when altering $\tau$. The major differences are found in the bumps during the day time from 6:00 to 18:00.  When a negative score on this component is computed, high traffic flow values relative too the central tendency must be present in the time spans  of the negative component loadings (e.g. it is the case for the time span between 12:00 and 20:00 when $\tau=0.05$).

\begin{figure}[hb!]	
	\centering
	\includegraphics[scale=0.3]{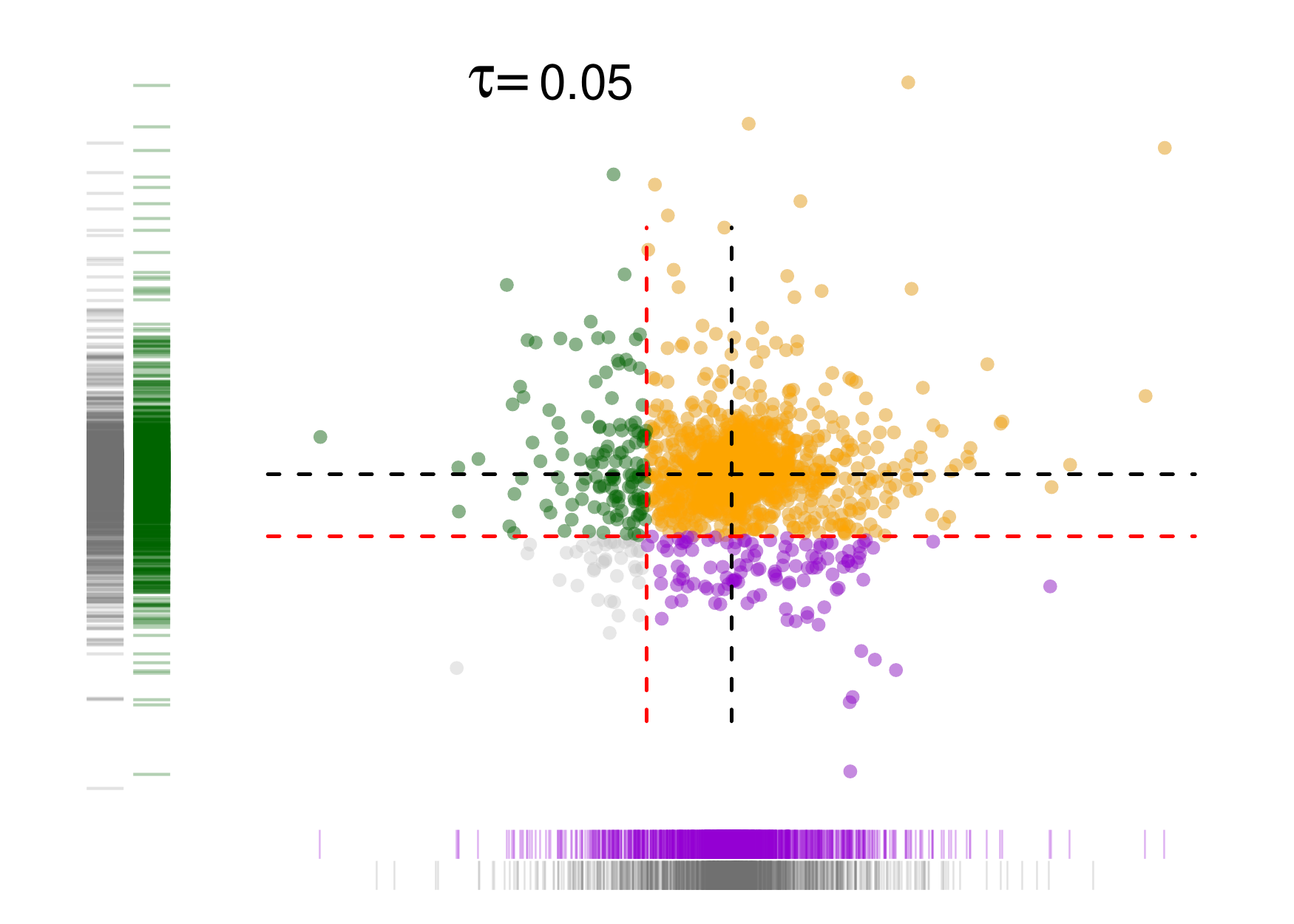}\includegraphics[scale=0.3]{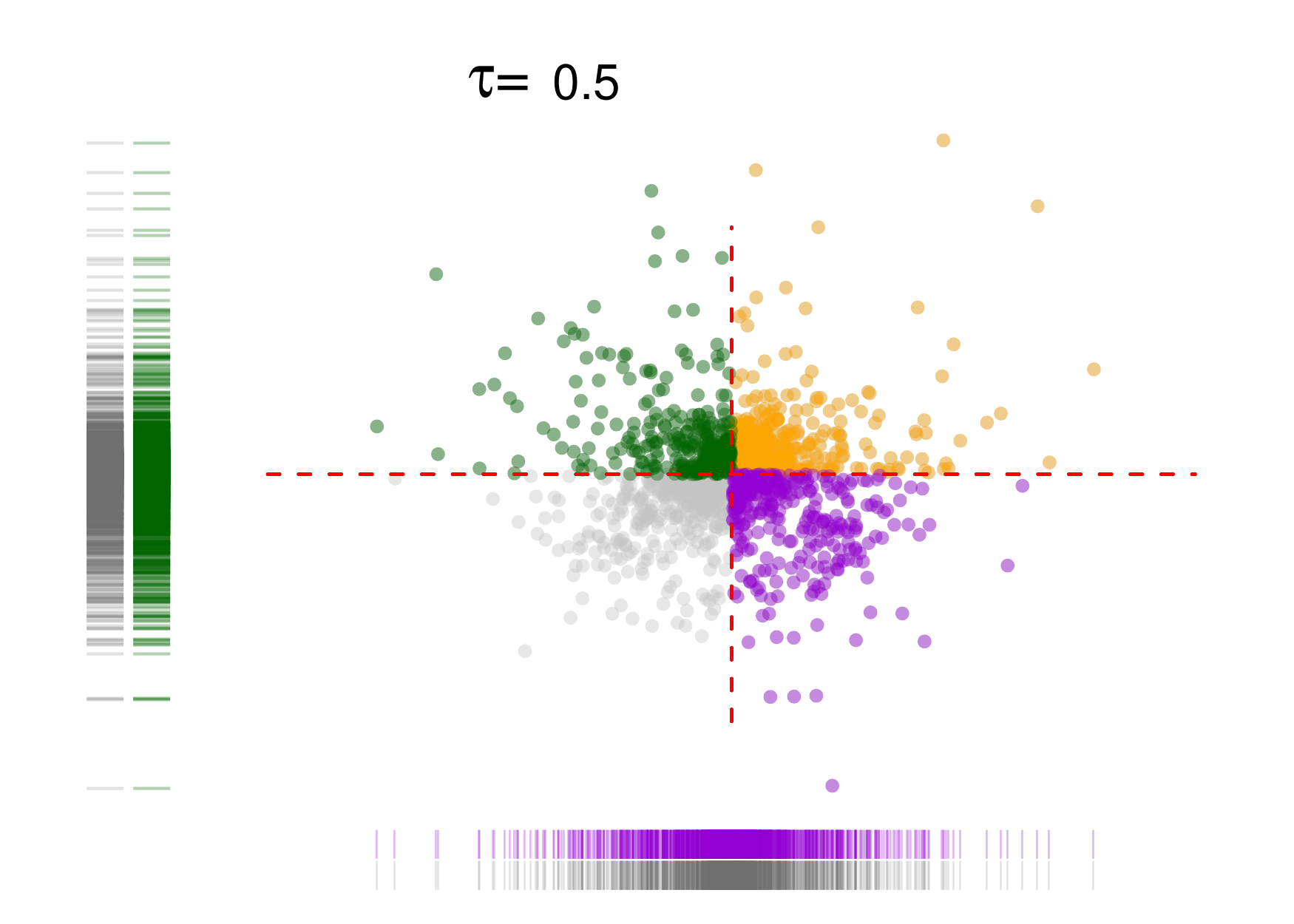}\includegraphics[scale=0.3]{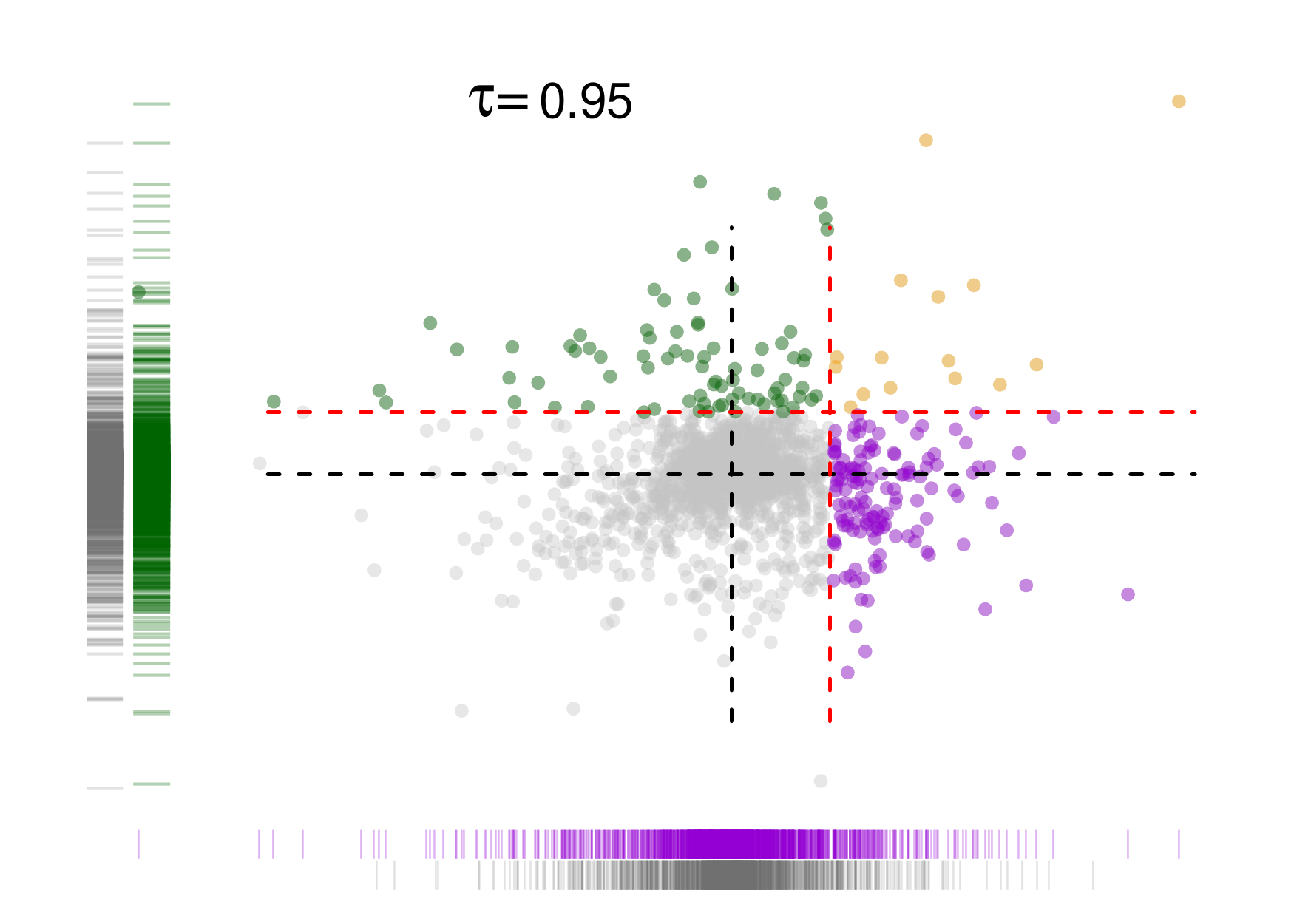}
	\caption{\scriptsize{Scores on the 3rd PEC (vertical axis) vs. scores on the 4th PEC (horizontal axis) for $\tau$ 0.05 (left), 0.5 (center), and 0.95 (right). The green (the purple) points indicate the profile scores, weighted with $\tau$ by the 3rd PEC (by the 4th PEC), the orange points represent the scores of the profiles weighted with $\tau$ by both components. The horizontal and vertical dashes visualize the projections on the 3rd (green) and on the 4th (purple) PECs. The gray dashes visualizing the projection on the 3rd and the 4th PCs ($\tau$=0.5) are added for better comparison.}}
		\label{points}	
\end{figure}

As noted before, the third and the fourth components exhibit some notable shape differences for varying $\tau$-levels. For instance, the fourth component has falling loadings around the noon and increasing loadings around the late rush hour as $\tau$ approaches 0.95. These shape corrections reflect the adjustments in the projection direction to the $\tau$-level shifting the center of our focus. In Fig. \ref{points}  the scores on the third PEC,  $\hat\phi_{\tau,3}^\top q_i$, $i=1,\ldots,n$ (vertical axis), versus the scores on the fourth PEC,  $\hat\phi_{\tau,4}^\top q_i$, $i=1,\ldots,n$ (horizontal axis) for $\tau$= 0.05, 0.5, and 0.95 are shown. The green color indicates the scores of the flow profiles weighted with $\tau$ by the third PEC (with other words, contained in $\mathcal I_{\tau,3}^+$); the purple color indicates the scores of the profiles weighted with $\tau$ by the fourth PEC  (contained in $\mathcal I_{\tau,4}^+$);  the orange points represent the scores of the  profiles weighted with $\tau$ by both components  (contained simultaneously  in $\mathcal I_{\tau,3}^+$ and in $\mathcal I_{\tau,4}^+$). We see higher spreads in the tails of the projections as the $\tau$-level departures from 0.5 (observe the dashed lines on horizontal and the vertical axes). 

As seen in Fig. (\ref{points}), the higher the level of $\tau$, the fewer traffic profiles are assigned to $\mathcal I_{\tau,k}^+$, and the more extreme their trajectories become with respect to the current projection direction. The memberships of traffic profiles in $\mathcal I_{\tau,k}^+ $ can be easily explored for discovering their directional extremes. When $\tau$ is far enough from 0.5, say 0.95, we directly label the profiles contained in $\mathcal I_{0.95,k} ^+$ as extreme in the direction $\hat\phi_{0.95,k}, k=1,\ldots,K$. When $\tau<0.5$, we focus on $\mathcal I_{\tau,k}^-$ as the profiles with larger weights are now contained in this set. The resulting labels are useful for determining whether external factors as weather or location influence the "extremeness" of the flow profile trajectory.
\begin{figure}[hb!]
	\centering
	\includegraphics[scale=0.89]{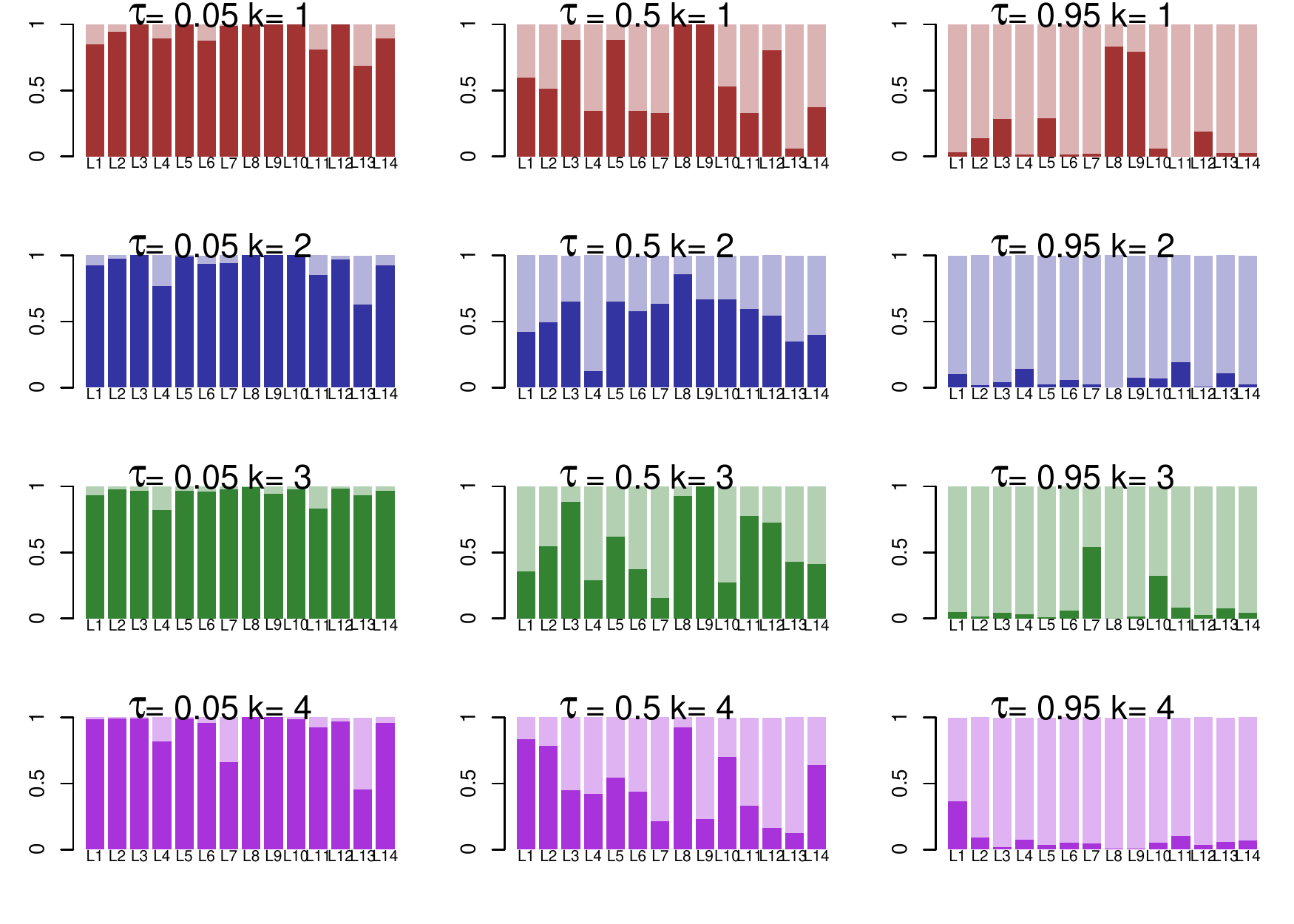}
	\caption{\scriptsize{The proportions of the days the traffic profiles of the locations $L1,\ldots, L14$ are assigned to $\mathcal I_{\tau,k} ^+$ (dark shade) or to $\mathcal I_{\tau,k} ^-$ (light shade) for $k=1,2,3,4$ (red, blue, green and purple respectively)) for different $\tau$-levels.}}
	\label{bars}
\end{figure}	
 
In Fig. (\ref{bars}), we show the resulting proportions of the days, the traffic profiles of the locations $L1,\ldots, L14$ are assigned to $\mathcal I_{\tau,k} ^+$ (dark shade) or to $\mathcal I_{\tau,k} ^-$ (light shade) for $k=1,2,3,4$ (red, blue, green and purple respectively)) for different $\tau$-levels. With $\tau$ far enough from 0.5, we are able to better grasp some 'problematic' points on the roundabout, which often produce extreme flow profiles in the sense of the corresponding component. For $\tau=0.95$, we observe some locations which enjoy the membership in $\mathcal I_{\tau,k} ^+$  notably more frequent than others. For instance, $\hat\phi_{\tau=0.95, k=4}$ assigns often the traffic profiles of $L1$ to its extreme category $\mathcal I_{\tau=0.95,k=4}^+$. The profiles in $L7$ and $L10$ are frequently assigned to $\mathcal I_{\tau=0.95,k=3}^+$, and those of $L8$ and $L9$ -- to $\mathcal I_{\tau=0.95,k=1}^+$.  When $\tau=0.05$, $L7$ is often chosen for $\mathcal I_{\tau=0.95,k=4}^-$, $L4$ for $\mathcal I_{\tau=0.95,k=2,3,4}^-$, and  $L13$ for $\mathcal I_{\tau=0.95,k=1,2,4}^-$. 

\begin{figure}[ht!]
	\centering
	\includegraphics[scale=0.9]{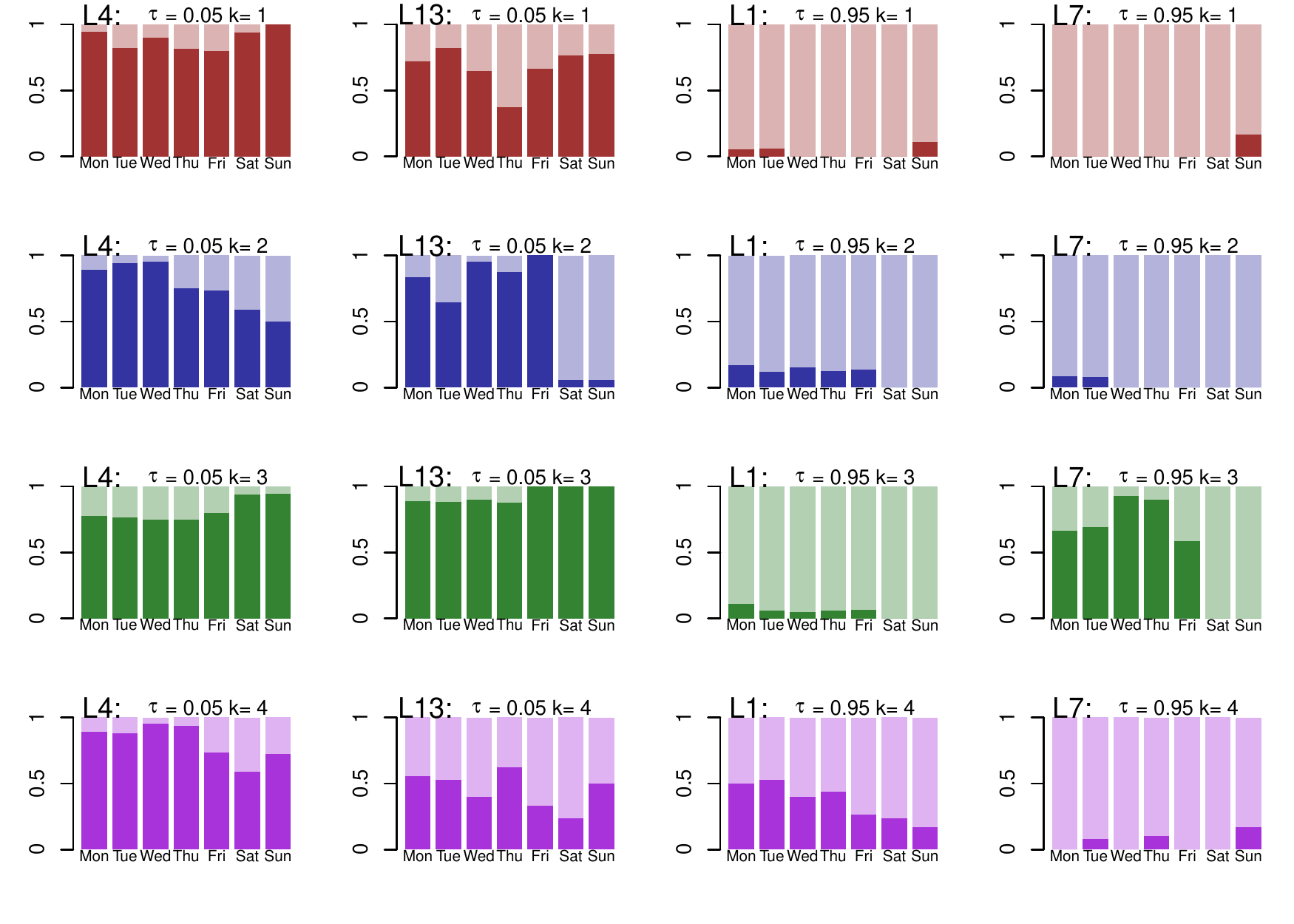}
	\caption{\scriptsize{
			The proportions of the days, the location specific daily profiles are in the set $\mathcal I_{\tau,k} ^+$ (dark shades)  and $\mathcal I_{\tau,k} ^-$ (light shades), $k=1,2,3,4$ (red, blue, green, purple)) for $L4$, $L13$ with $\tau$=0.05 (1st and 2nd columns) and for $L1$, $L7$ with $\tau=$0.95 (3rd and 4th columns) against the day type.
		}}	
		\label{bars_day}
\end{figure}

To better understand the circumstances under which the directional extremes happen in our considered locations, one can enclose other sources of information as the day type and compare the profile membership proportion in $\mathcal I_{\tau,k}^+$ with respect to the day type. In Fig. (\ref{bars_day}),  we show the proportion of days the daily location profile was a member of $\mathcal I_{\tau,k} ^+$ (dark shades) and of $\mathcal I_{\tau,k} ^-$ (light shades) for $k=1,2,3$ and 4 (red, blue, green and purple). As before,  when $\tau>0.5$ the relevant set is $\mathcal I_{\tau,k} ^+$, whereas $\mathcal I_{\tau,k} ^-$ is worth attention when $\tau<0.5$. From Fig. (\ref{bars_day}), we can infer whether the day type is essential for the membership in the respective sets. For instance, the directional extremes in $L4$ with respect to $\hat\phi_{0.05,k=3}$ seem to occur mainly on the weekdays from Monday to Friday, whereas the extremes with respect to $ \hat\phi_{0.05,k=4}$ prevail on the weekends from Friday to Sunday (the two last panels in the left column of Fig. (\ref{bars_day})). 

\begin{figure}[ht!]	
	\centering
	\includegraphics[scale=0.9]{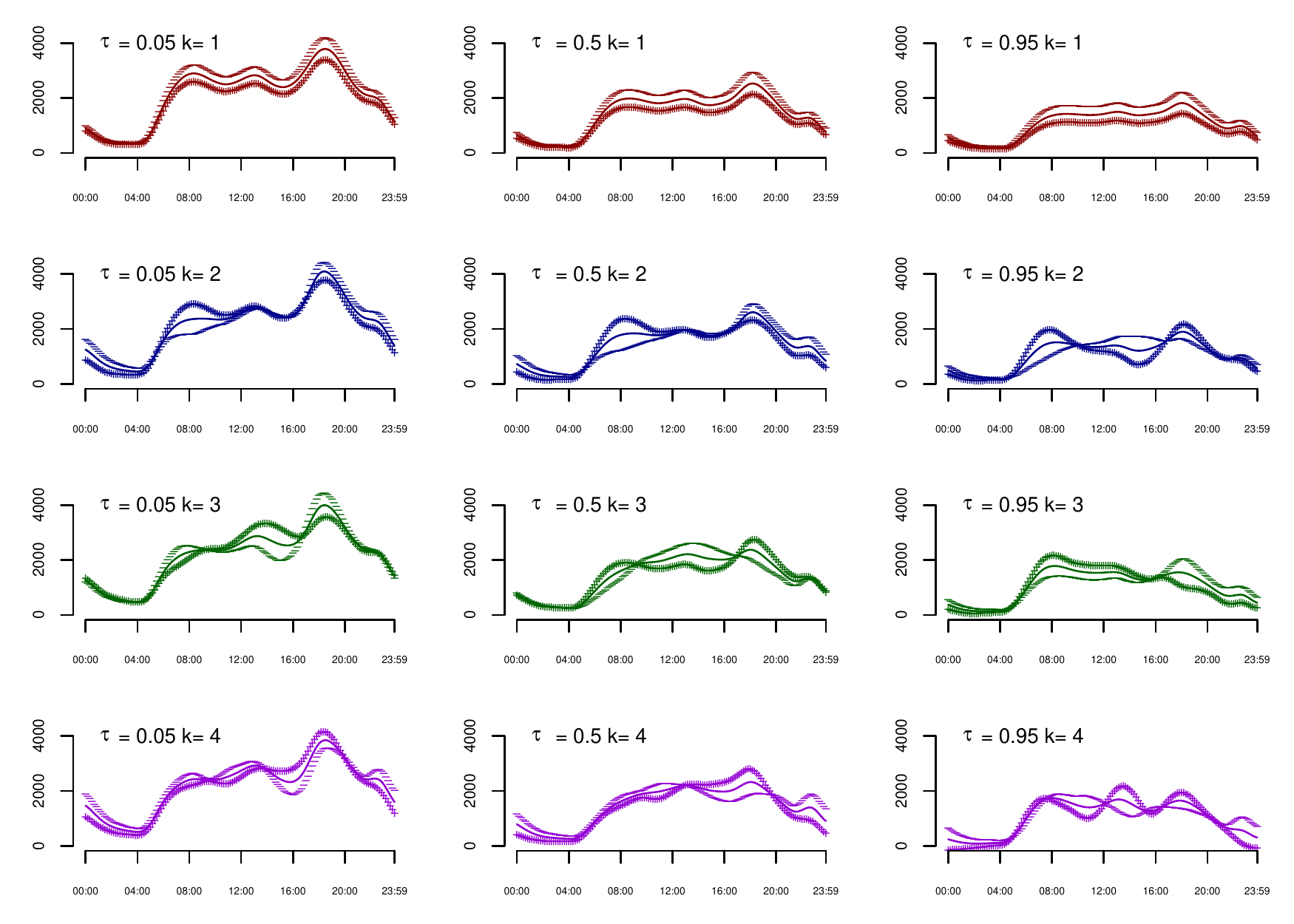}
	\caption{\scriptsize{Generic flow profiles, resulting from adding(++) / subtracting (- -) a scaled multiple of the first (top), second (top middle), third (bottom middle), and fourth (bottom) components to/ from the average profile for $\tau$ = 0.05 (left), 0.5 (center), and 0.95 (right).
	}}		
\label{pec_plmin}
\end{figure}

To gain a deeper insight on what exactly distinguishes these directional extreme daily profiles in $\mathcal I_{\tau,k} ^+$ or $\mathcal I_{\tau,k} ^-$, we look at
the effects, resulting from adding or subtracting a suitable multiple of the $k$th component to the $\tau$ anticipated central tendency $\tilde e_{\tau, k}$.
Fig. (\ref{pec_plmin}) depicts the plus-minus-component-multiple effects for $\tau=0.05,0.5, 0.95$. In the central column, the depicted PECs correspond to the classical PCs. In the left and the right columns, we observe the plus-minus-component-multiple effects of $\phi_{\tau=0.05,k}$ and $\phi_{\tau=0.95,k}, k=1, \ldots,4$. Whereas for $\tau=0.05$ the PECs identify the traffic profiles with generally higher traffic flow levels ($k=1$) accentuating the afternoon peak ($k=2,3,4$), setting $\tau=0.95$ changes the focus to the profiles with rather lower flow values ($k=1$) highlighting the deviations  also the early rush hour ($k=2,3$) and the midday flow ($k=4$). 

\begin{figure}[ht!]
	\centering
	\includegraphics[scale=0.9]{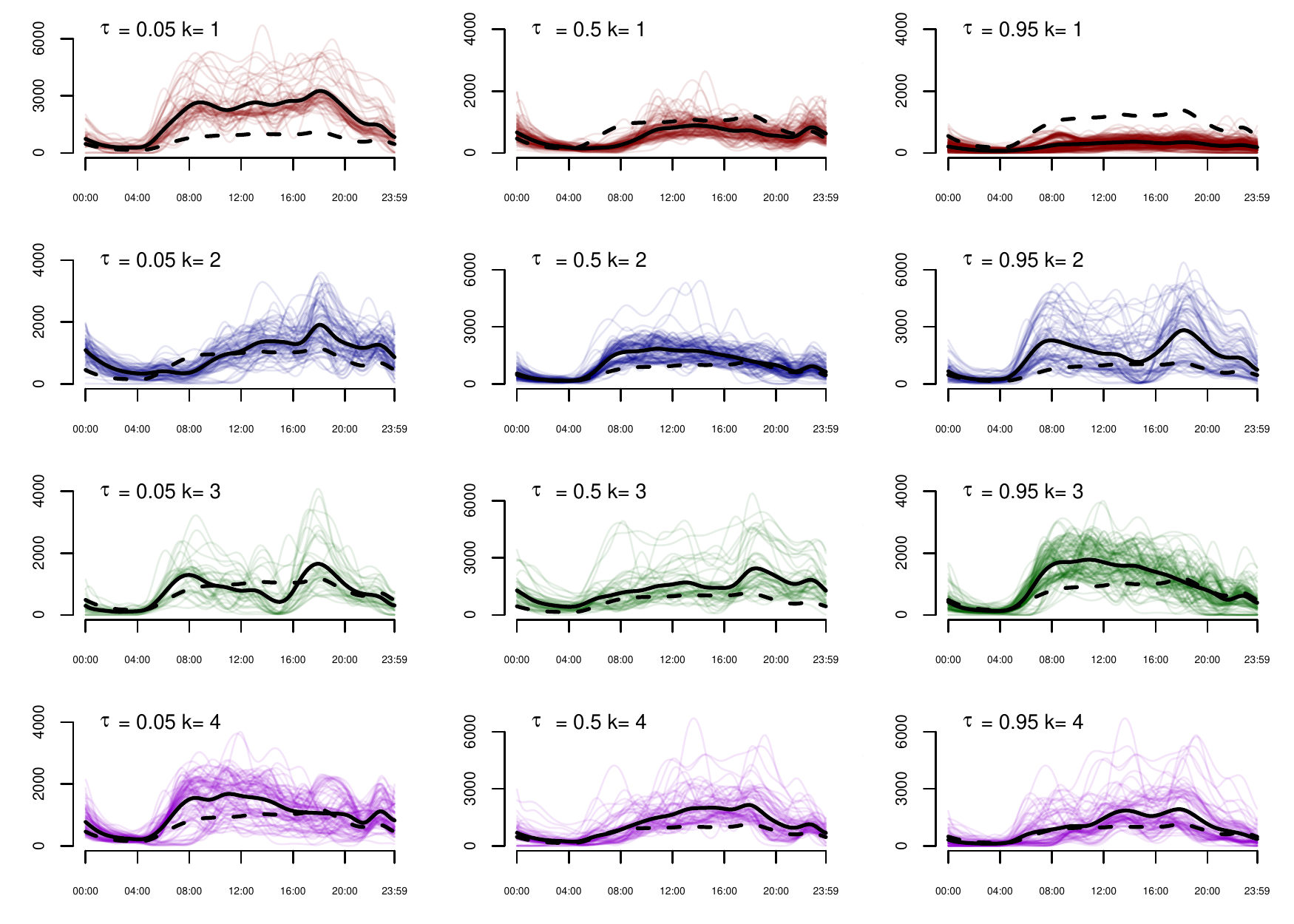}
	\caption{\scriptsize{
			The traffic flow profiles contained in $\mathcal I_{\tau=0.05,k} ^-$ (the left column), $\mathcal I_{\tau=0.5,k} ^+$ (the middle column), and $\mathcal I_{\tau=0.95,k} ^+$ (the right column) for $k=1,\ldots 4$ (top-bottom ordering, red, blue, green, purple lines respectively), their average shape (black solid line) as opposed to the average shape of the daily profiles not contained in the respective sets (black dashed line).
		}}	
		\label{paths}
\end{figure}

The profiles in $\mathcal I_{\tau=0.05,k}^-$, in $\mathcal I_{\tau=0.5,k}^+$, and in $\mathcal I_{\tau=0.95,k}^+$ are plotted in Fig. (\ref{paths}) against the average of the reciprocal sets. We see that the respective directional extreme profiles for $\tau=0.05$ and for $\tau=0.95$ definitely show exceptional trajectories compared to the average of the others.
For $\tau=0.05$ the red trajectories ($k=1$), as expected,  exhibit higher than average flow values, for $\tau=0.95$ - the vice versa is the case. The profiles contained in $\mathcal I_{\tau=0.05,k=2,3,4} ^-$ and in $\mathcal I_{\tau=0.95,k=2,3,4} ^+$ (the left and the right columns of Fig. (\ref{paths})) inherit the plus-minus-component-multiple effects of the respective PECs from Fig. (\ref{pec_plmin}). 

One can thus use the panels of Fig. (\ref{bars}) and Fig. (\ref{pec_plmin}) for $\tau$ far from 0.5 to extract the profiles with the desirable directional extreme behavior. This framework naturally enables simple identification and subsequent exploration of the obtained directional traffic profiles extremes. %and provides new insights in the structure of daily traffic flow profiles for further exploitation. 

\section{Conclusion}
A novel dimension reduction technique called principal component analysis in an asymmetric norm is applied in this paper to study multivariate extremes of traffic flow data. 
Our approach adopts a directional definiton of extremeness and goes beyond the classical principal component introducing its asymmetric version directed by a prespecified expectile level $\tau$ towards the extremes. The resulting principal expectile component points in the direction of the largest $\tau$-variance of the projection. At the core is the idea of asymmetric weighting of the data points with either $\tau$ or $1-\tau$.
A step-by-step alternating computing algorithm is presented for obtaining the principal expectile components from the data. 

Using the daily traffic flow profiles collected from January to July 2018 with 14 detectors in different points of the roundabout on Ernst-Reuter-Platz in the city center of Berlin, Germany, we compute the first four sample principal expectile components for different $\tau$-levels. The components project the flow profiles to lower dimensions in such a way, that the asymmetrically weighted variation around the sample expectile of the projection is the largest possible. The obtained low dimensional representation of the profile shapes focuses on their extreme trajectories in the directional sense determined by the loadings of the computed principal expectile components.

Our subsequent analysis demonstrates, that the computed principal expectile components reveal some interesting structures in the directional extremes.  Since all flow profiles in the collection are weighted with either $\tau$ or $1-\tau$ in each principal expectile direction, the set of the high-weighted flow profiles can be easily used for identification and exploration of  the corresponding directional extremes. As shown, the extracted information facilitates selection of the flow profiles with particular shapes, location- and day-related specifics. 

The proposed approach naturally extends PCA-based methodological basis for dimension reduction and pattern recognition in traffic flow profiles. Its potential in improving performance of predictive models and enhancing control applications, that include dimension reduction, should be explored in future work. Further topics worth investigating are (currently absent) criteria for choosing an appropriate number of principal expectile components and strategies for choosing the most informative $\tau$-level.

% Double-space the bibliography

\end{document}